\documentclass[preprint,floatfix
,aps,prd,superscriptaddress]{revtex4}

\usepackage[utf8]{inputenc}
\usepackage{graphicx}
\usepackage{tikz}
\usepackage{amssymb,amsmath,mathtools,comment}
\newcommand{\ket}[1]{|#1\rangle}
\newcommand{\bra}[1]{\langle #1|}
\newcommand{\lrp}[1]{\left( #1 \right)}
\newcommand{\lrb}[1]{\left[ #1 \right]}
\newcommand{\stackh}{\stackrel{\mathcal{(H) }}{\sim}}
\begin{document}

\title{
Near-horizon aspects of acceleration radiation by free fall of an atom into a black hole}

\author{H. E. Camblong}
\affiliation{Department of Physics and Astronomy, University of San Francisco. San Francisco, California 94117-1080, USA}
\author{A. Chakraborty}
\affiliation{Department of Physics, University of Houston. Houston, Texas 77024-5005, USA}
\author{C. R. Ord\'{o}\~{n}ez}
\affiliation{Department of Physics, University of Houston. Houston, Texas 77024-5005, USA}
\affiliation{Departmento de F\'{i}sica, Facultad de Ciencias Naturales y Exactas, Universidad de Panam\'a, Rep\'ublica de Panam\'a.}
\affiliation{Department of Physics and Astronomy, Rice University, MS 61, 6100 Main St., Houston, TX 77005.\footnote{Current address}}

\date{\today}
\begin{abstract}
A two-level atom freely falling towards a Schwarzschild black hole was recently shown to detect radiation in the Boulware vacuum in an insightful paper [M. O. Scully et al.\/, PNAS 115(32), 8131 (2018)]. The two-state atom acts as a dipole detector and its interaction with the field can be modeled using a quantum optics approach. The relative acceleration between the scalar field and the detector causes the atom to detect the radiation. In this paper, we show that this acceleration radiation is driven by the near-horizon physics of the black hole. This insight reinforces the relevance of near-horizon conformal quantum mechanics for all the physics associated with the thermodynamic properties of the black hole. We additionally highlight the conformal aspects of the radiation that is given by a Planck distribution with the Hawking temperature.
\end{abstract}


\maketitle


\section{Introduction}
\label{sec:intro}
\noindent 
A black hole is a spacetime region, derived from a classical solution to the Einstein field equations of general relativity, from which no physical signals can escape. However, Hawking's seminal works \cite{hawking1,hawking2,hawking3} showed that black holes can radiate, as was soon corroborated by a series of papers by Unruh, Davies, Fulling, and many others \cite{unruh76,ufd76,davies78,df77,dewitt79,wald75,parker75}. It was shown that this radiation is caused by quantum effects in curved spacetime and is closely related to black hole thermodynamics.
Almost half a century later, the thermal nature of black hole radiation has been extensively studied, but we do not yet have a final picture of the Hawking effect and all of its implications. One aspect of particular interest is the role of the observer in the detection of this thermal radiation. Unruh and Wald \cite{unruhwald84} showed that an accelerated observer experiences particles in a thermal bath in the Minkowski vacuum of an inertial observer. The detection of particles in the inertial vacuum by an accelerated detector is known as the Unruh effect and is closely related to the Hawking radiation from a black hole \cite{unruh76,unruhreview}. A more recent development in understanding the Unruh effect is the use of quantum optics to model the accelerated detector by a two-state atom \cite{belyaninscully03,belyaninetal}. This model was applied by Scully et al.\ in a more recent thought-provoking paper \cite{scully2018} to show that an atom freely falling through a Boulware vacuum \cite{boulware} of a Schwarzschild black hole experiences thermal radiation. At a first glance, this seems to violate the equivalence principle since the freely falling atom is in a locally inertial frame. However, it is the relative acceleration between the field modes (defined with boundary conditions at asymptotic infinity) and the freely falling atom that gives rise to the acceleration radiation. This was subsequently illustrated by a series of gedanken experiments designed by Fulling \cite{fulling2018}.
    
Another insightful approach to black hole thermodynamics is based on the conformal symmetry near the event horizon of a black hole. The relation of the central charge of the Virasoro algebra in the backdrop of conformal field theory with the black hole entropy was discussed in several papers \cite{carlip1,carlip2,birminghamsen,cardy}. One perspective involves finding the connection between conformal quantum mechanics (CQM), which is essentially conformal field theory in 0+1 dimension \cite{dffCQM}, and the Bekenstein-Hawking entropy \cite{guptasen,nhcamblong,nhcamblong-sc,moretti,vaidya,tightnesscamblong}. In Refs.~\cite{nhcamblong,nhcamblong-sc}, black hole thermodynamics was shown to emerge from CQM as the near-horizon approximation to the field modes, leading to the Bekenstein-Hawking entropy \cite{bekenstein} interpreted via a brick-wall model~\cite{thooft}, with a natural cutoff of the order of the Planck-length scale. 

In this paper, we make an explicit connection between the near-horizon CQM framework developed in Ref.~\cite{nhcamblong} and the quantum optics approach advanced by Scully et al.~\cite{scully2018}. We show explicitly that the main contribution to the excitation probability of the freely falling atom described in Ref.~\cite{scully2018} comes from the scale-invariant behavior of the near-horizon CQM field modes. In its final form, this result exhibits a leading near-horizon radiation governed by conformal invariance and given by a Planck distribution with the Hawking temperature. Therefore, these findings further confirm that the radiation emitted by the freely falling atom is a near-horizon conformal phenomenon.

Moreover, the systematic application of the near-horizon expansion defined in Refs.~\cite{nhcamblong,nhcamblong-sc} allows us to extend the analytical calculation of the excitation probability to the whole class of D-dimensional generalized Schwarzschild metrics with general initial conditions for the free motion of the atom. We thus show that the final result is independent of these generalizations, governed by conformal invariance, and with details matching the special case considered in Ref.~\cite{scully2018}.  

This paper is organized as follows. In Sec.~\ref{sec:background}, we briefly discuss the background needed for the near-horizon treatment of the problem. This section is divided into two parts describing the emergence of the CQM equation from the Klein-Gordon equation, and the basic tools of quantum optics needed for the subsequent calculations. In Sec.~\ref{sec:nhqoapproach}, we use the near-horizon behavior to extend the quantum optics formalism to a more general setting, viz., generalized Schwarzschild metric with arbitrary initial conditions. In Sec.~\ref{sec:conformal_aspects}, we further highlight the consequences of the near-horizon conformal symmetry. The paper concludes in Sec.~\ref{sec:Discussion} with a brief discussion on the implications and possible applications of these results. Finally, in the appendix, we provide some technical details related to the results discussed in the main text.

\section{Background \label{sec:background}} 
\subsection{Near-horizon CQM equation for generalized Schwarzschild metric\label{subsec:CQMderivation}}
\noindent Throughout the paper we will adopt natural units (unless stated otherwise), with $\hbar =1$ and $c=1$, in conjunction with the metric conventions of Ref.~\cite{MTW-gravitation}.
We consider the family of static and spherically symmetric spacetime geometries, which are described by the generalized Schwarzschild metric
\begin{equation}
ds^{2}=- f (r) \,  dt^{2}+\left[ f(r) \right]^{-1} \, dr^{2}+ r^{2} \, d \Omega^{2}_{(D-2)}\; ,
\label{eq:RN_metric}
\end{equation}
in $D$ spacetime dimensions (with $D \geq 4$), where $d \Omega^{2}_{(D-2)}$ stands for the metric on the unit $(D-2)$-sphere, $S^{D-2}$, that foliate the spacetime manifold. This class of metrics extends the familiar 4D Schwarzschild solution to $D$ dimensions, and also includes the D-dimensional Reissner-Nordstr\"{o}m (RN)
metric~\cite{mye:86}, as well as combinations of these with a cosmological constant, and black hole solutions with additional charges~\cite{ortin}. The near-horizon analysis will be centered on the functional dependence of the fields in the neighborhood of the outer event horizon at $r=r_{+}$, employing the particular set of generalized Schwarzschild coordinates $(t,r, \Omega)$. We briefly review the setup developed in \cite{nhcamblong,nhcamblong-sc}, in which this singular-coordinate choice (around a coordinate singularity) displays the conformal quantum-mechanical symmetry from the outset and gives additional insight into black hole thermodynamic relations. The full derivation has been shown in Appendix~\ref{app:cqm} for completeness. We start with the Euler-Lagrange equation satisfied by the scalar field in the black hole gravitational background, which is given by
\begin{equation}
\left[ \Box - \left(m^{2} + \xi R\right) \right] \Phi \equiv \frac{1}{ \sqrt{-g} }\partial_{\mu} \left(\sqrt{-g} \,g^{\mu \nu}\,\partial_{\nu} \Phi\right)- \left(m^{2} + \xi R\right)\Phi= 0\; .
\label{eq:Klein_Gordon_basic}
\end{equation}
This is the Klein-Gordon equation in curved spacetime. For the class of metrics~(\ref{eq:RN_metric}), we can consider the following mode expansion of the scalar field
\begin{equation}
\Phi(t,r,\Omega) = \sum_{n,l,m} \lrb{a_{nlm} \, \phi_{nlm}(r,\Omega,t) + H.c.} \; , \label{eq:quantization}
\end{equation}
where $a_{nlm}$ is the field annihilation operator, H.c.\ means hermitian conjugate, and $\phi_{nlm}$ constitute a complete set of orthonormal solutions to Eq.~(\ref{eq:Klein_Gordon_basic}) with respect to the corresponding Klein-Gordon inner product~\cite{DeWitt_QFT-global}. The use of Schwarzschild coordinates selects these particular modes for the expansion of Eq.~(\ref{eq:quantization}); and the corresponding Killing time $t$ leads to a definition of the positive frequency choice from which a Fock space with mode occupation numbers is constructed. This includes the existence of an associated Boulware vacuum $ \ket{0_{B}}$  such that~\cite{boulware,DeWitt_QFT-global}
\begin{equation}
a_{nlm}  \ket{0_{B}} = 0
\; 
\label{eq:Boulware-vacuum}
\end{equation}
for all modes, and which asymptotically behaves as the Minkowski vacuum at infinity.

Equation~(\ref{eq:Klein_Gordon_basic}) is separable in Schwarzschild coordinates with the following ansatz
\begin{equation}
\phi_{nlm} (t,r, \Omega) = \chi (r)\, u_{nl}(r)Y_{lm}(\Omega)e^{-i\omega_{nl}t}\; ,
\label{eq:separation_of_variables}
\end{equation}
where the angular part is given by the ultra-spherical harmonics $Y_{lm}(\Omega)$ and the time dependence involves frequencies $\omega_{nl}$. In addition, the particular choice of $
\chi (r)= [f(r)]^{-1/2} \,r^{-(D-2)/2}$ reduces the radial part of the Klein-Gordon equation to its normal form 
\begin{equation}
u_{nl}''(r) +I_{(D)} (r; \omega_{nl}, \alpha_{l,D}  )\,u_{nl}(r) =0\;  ,
\label{eq:Klein_Gordon_normal_radial}
\end{equation}
where $I_{(D)}$ is an effective potential whose full form is given in Appendix~\ref{app:cqm}. The behavior of the modes arising from Eq.~(\ref{eq:Klein_Gordon_normal_radial}) can be examined near the outer horizon ${\mathcal H}$, $ r \sim r_{+}$, with $r=r_{+}$ being the largest root of the scale-factor equation $f(r)=0$. This can be performed by the shifted variable $x= r -r_{+}$, in terms of which the Taylor series for the scale factor $f(r)$ starts at first or higher orders. In this paper, we only consider {\em nonextremal\/} metrics that satisfy the condition $f'_{+} \equiv f'(r_{+}) \neq 0$. Then, the expansions of $f(r)$ and its derivatives are given by 
\begin{eqnarray}
f(r) & \stackrel{(\mathcal H)}{\sim} & f'_{+}  \, x \left[ 1 + \mathcal{O}(x) \right]\; ,\nonumber\\
f'(r) & \stackrel{(\mathcal H)}{\sim} & f'_{+} \left[ 1 + \mathcal{O}(x) \right]\; ,\nonumber\\
f''(r) & \stackrel{(\mathcal H)}{\sim} & f''_{+} \left[ 1 + \mathcal{O}(x) \right]\; ,
\label{eq:nh-expansions}
\end{eqnarray}
where $f''_{+} \equiv f''(r_{+}) $ and the notation $\stackrel{(\mathcal H)}{\sim}$ will be used to represent the hierarchical expansion about the horizon. \\
With this near-horizon expansion, the effective potential in Eq.~(\ref{eq:Klein_Gordon_normal_radial}) can be simplified significantly and, up to the leading-order term in $x$, is given by the form (as shown in Appendix~\ref{app:cqm})
\begin{equation}
u''(x)+\frac{ \lambda_{\rm eff} }{x^{2}}\,\left[ 1 + \mathcal{O}(x) \right]u (x)=0\;  ,
\label{eq:Klein_Gordon_conformal}
\end{equation}  
where, by abuse of notation, we have replaced $u(r)$ by $u(x)$. Equation~(\ref{eq:Klein_Gordon_conformal}) indicates that dominant physics near the horizon is driven by the interaction
\begin{equation}
V_{\rm eff} (x) = - \frac{ \lambda_{\rm eff} }{x^{2}}\, ,  \; \; \; \; \lambda_{\rm eff} = \frac{1}{4} + \Theta^{2}\, , \; \; \; \; \Theta= \frac{\omega}{  f'_{+} } \; ,
\label{eq:conformal_interaction}
\end{equation}  
which corresponds to a one-dimensional effective Hamiltonian $H= {p}_{x}^{2}-  \lambda /x^{2}$. This is the well-known long-range representative of conformal quantum mechanics~\cite{camblongmp}. Thus, our derivation shows that the near-horizon physics exhibits an {\em asymptotic conformal symmetry\/}.

\subsection{Acceleration radiation by an atom falling freely towards a black hole \label{subsec:QObackground}}
\noindent In this subsection we use the setup described in Ref.~\cite{scully2018}. A two-level dipole atom, which acts as the detector, falls freely towards the black hole described by the generalized Schwarzschild metric (\ref{eq:RN_metric}).  Our goal is to probe the atom's acceleration radiation (Unruh effect). The Boulware vacuum state~\cite{scully2019,brick-wall}, defined by Eq.~(\ref{eq:Boulware-vacuum}), allows us to single out this form of radiation and explicitly separate it from the one due to the black hole itself, i.e., the Hawking effect~\cite{Barbado_HvsU}. Specifically, while in this Boulware state $ \ket{0_{B}} $ there is no Hawking radiation, a freely falling observer or detector will perceive particles, as we explicitly show below. 
%


As the atom falls towards the black hole, it will detect radiation by going to the excited state and emitting a photon. The probability of this process can be expressed as
\begin{equation}
P_{exc} = \frac{1}{\hbar^2}\left|\int d\tau \;\bra{1_{\bf n},a}V_I(\tau)\ket{0,b}\right|^2 \label{eq:P_ex_expression}
\end{equation}
where $\ket{b}$ and $\ket{a}$ are respectively the ground and the excited state of the atom and $\tau$ is the atom's proper time. In addition, $\ket{1_{\bf n}}$ represents the one-photon mode with quantum numbers ${\bf n}$ and $\ket{0} \equiv \ket{0_{B}}$ denotes the Boulware vacuum state of the field. 

The relevant interaction potential $V_I(\tau)$ needed for Eq.~(\ref{eq:P_ex_expression}) is given by the quantized dipole interaction 
\begin{equation}
V_I(\tau) = \hbar g \left[a_{\bf n} \phi_{\bf n}(r (\tau),t(\tau)) + H.c.\right]\lrp{\sigma_- e^{-i\nu \tau} + H.c.} \, ,
\label{QO_interaction_potential}
\end{equation}
where $\sigma_-$ is the lowering operator for the atom, $\nu$ is the atom frequency, and the coupling constant $g$ denotes the strength of the interaction, which will be assumed to be weak in the derivations of this paper. In addition, in Eq.~(\ref{QO_interaction_potential}), $\phi_{\bf n}(r,t)$ are the field modes with quantum numbers ${\bf n}$, which can be obtained from the Klein-Gordon equation as in Subsec.~\ref{subsec:CQMderivation}, and $a_{\bf n}$ stands for the associated annihilation operators. For our choice of generalized Schwarzschild coordinates, as assumed in the expansion of the field modes, Eq.~(\ref{eq:quantization}), the quantum numbers are explicitly ${\bf n}\equiv (n,l,m)$. As displayed in Eq.~(\ref{QO_interaction_potential}) and in the treatment that follows below, we will assume that we can neglect the angular dependence of the modes. (We will either consider the simplest transitions from the ground state to an excited state without angular momentum, or will assume a near horizon approximation where such dependence would only yield a phase factor not affecting the probability.)

Several important remarks are in order. First, Eq.~(\ref{QO_interaction_potential}) models a dipole interaction Hamiltonian whose scale is given by the coupling $g = \mu E/\hbar$, where $\mu$ is the atomic dipole moment and $E$ is the electric field; thus, $g$ has dimensions of frequency,  inverse length, and mass in natural units
$c=1$, $\hbar =1$. Second, the remainder of the expression in Eq.~(\ref{QO_interaction_potential}) involves the field and atom operators, with all factors being dimensionless to guarantee that the overall Hamiltonian also has 
dimensions of frequency or mass. Third, this Hamiltonian describes a simplified model where ordinary vector (spin-1) photons are replaced by scalar (spin-0) ``photons.'' Fourth, the normalization of the field modes $\phi$ is somewhat arbitrary, and needs to be specified consistently with the dimensionless requirement. The corresponding normalization can be achieved by including all relevant factors with coordinate dependence and subsuming them into a pure phase function---this is typically straightforward, as it corresponds to the local outgoing/ingoing waves that can be defined around coordinate singularities (e.g., the event horizon); see Sec.~\ref{sec:nhqoapproach}.

It is clear from the expression of $P_{exc}$ in Eq.~(\ref{eq:P_ex_expression}) that only the term corresponding to $a_{\bf n}^\dagger\sigma_-^\dagger$ will give non-zero contribution to the probability. This enables us to write Eq.~(\ref{eq:P_ex_expression}) in the following more explicit form
\begin{equation}
P_{exc} = g^2 \left|\int\; d\tau\; \phi^*(r (\tau),t(\tau)) e^{i\nu\tau}\right|^2. \label{eq:P_ex_explicit}
\end{equation}
In order to find the detection probability $P_{exc}$, we need the expression for the field modes and the trajectory of the atom in free fall which is described in the next section.

\section{Near horizon description of the acceleration radiation by a freely falling atom \label{sec:nhqoapproach}}
The field modes required to calculate the excitation probability $P_{exc}$ in Eq.~(\ref{eq:P_ex_explicit}) can be obtained using the CQM equation (\ref{eq:Klein_Gordon_conformal}), which gives a fundamental pair of near-horizon outgoing/ingoing waves; in particular, we will select the outgoing wave for the radiation emitted outwards by the atom from the neighborhood of the event horizon. This is given by
\begin{equation}
    u(x) = x^{\frac{1}{2}+\sqrt{\frac{1}{4}-\lambda}} = \sqrt{x} \, x^{i\Theta} 
\end{equation}
where $\Theta = {\omega}/{f_+'}$ as defined in Eq.~(\ref{eq:conformal_interaction}). We thus combine all the factors together to get the field mode $\phi(r,t)=\chi(r)u(r)e^{-i\omega t}$, where we will assume (as in Subsec.~\ref{subsec:QObackground}) that the angular dependence is not needed. Then, in the near-horizon expansion,
\begin{equation}
    \chi(r) = [f(r)]^{-1/2}r^{-(D-2)/2} \stackrel{(\mathcal{H})}{\sim}\; \frac{1}{\sqrt{x}\sqrt{f_+'}}(r_+)^{-(D-2)/2}(1+\mathcal{O}(x)) \; .
    \end{equation}
Therefore,
    \begin{equation}
    \phi(r,t) \stackh\; \frac{1}{\sqrt{f_+'}} \, r_+^{-(D-2)/2} \; x^{i\Theta} e^{-i\omega t} 
    \leadsto 
        \phi(r,t) \stackh\;  x^{i\Theta} e^{-i\omega t}  =   e^{-i\omega ( t - \ln x/f_+')} \;.
    \label{eq:conformal_field_modes}
\end{equation}
In the last step of Eq.~(\ref{eq:conformal_field_modes}),
 a pure-phase outgoing wave in the neighborhood of the event horizon is extracted, by removing the extra 
 constant factors (such as $r_+$ and $f_+'$) in the leading near-horizon approximation. (Incidentally, this can also be done most efficiently with semiclassical WKB techniques~\cite{camblongmp}.) Alternatively, this identification is equivalent to using the near-horizon expansion of the Klein-Gordon equation in Eddington-Finkelstein coordinates as shown in Appendix~\ref{app:efconformal}.

The spacetime trajectories for free-fall motion of the atoms in a background metric $g_{\mu\nu}$ are described by the geodesic equations. For a static and spherically symmetric metric defined by Eq.~(\ref{eq:RN_metric}), there is invariance under time translations and invariance under spatial rotations involving $(D-1)(D-2)/2$ planes. These symmetries lead to their associated conserved energy and components of the angular momentum tensor, and a corresponding number of Killing vectors. For $D=4$, the latter reduce to the familiar 3 components of angular momentum. All of the angular momentum components but one can be fixed to define a single plane for the orbit where an azimuthal angle $\phi$ can be used. This procedure reduces the problem to finding the geodesics with initial conditions defined by two conserved quantities
\begin{equation}
e = - \boldsymbol\xi \cdot {\bf u} = f (r) \frac{dt}{d\tau}
\; \; \;  , \; \; \; 
\ell = \boldsymbol\eta \cdot {\bf u} = r^2\frac{d\phi}{d\tau} \; ,
\label{eq:conserved-quantities}
\end{equation}
in terms of the Killing vectors $  \boldsymbol\xi = \partial_{t}$ and $\boldsymbol\eta = \partial_{\phi}$ and spacetime velocity $ {\bf u}$. More precisely, these are the energy per unit mass $e\equiv {E}/m$, and angular momentum per unit mass $\ell\equiv{L}/m$ (in terms of the mass $m$ of the atom). For a free fall from a fiducial point, with initial specific energy $e$ and initial specific angular momentum $\ell$, these conserved quantities give the initial conditions. When the Killing symmetries are enforced, the first-order form of the geodesic equations for timelike geodesics become~\cite{MTW-gravitation}
\begin{align}
    \frac{dt}{d\tau} &= \frac{e}{f(r)}\;.\label{eq:geodesic_t} \\
      \frac{dr}{d\tau} &= -\sqrt{e^2- f(r)\lrp{1+\frac{\ell^2}{r^2}}}\;,\label{eq:geodesic_tau}\\
   \frac{d\phi}{d\tau}  & = \frac{\ell}{ r^2} \; \label{eq:geodesic_phi}
\end{align}
(where Eq.~(\ref{eq:geodesic_tau}) represents velocity normalization combined with the other first integrals of the motion~(\ref{eq:conserved-quantities})). In particular, the negative sign in Eq.~(\ref{eq:geodesic_tau}) indicates the in-falling motion of the atom. We can integrate these equations to get the atom's proper time $\tau$ and the Schwarzschild coordinate time $t$ in terms of the radial variable $r$, 
\begin{align}
    \tau &= - \int_{r_0}^{r} \frac{dr}{\sqrt{e^2- f(r)\lrp{1+\frac{\ell^2}{r^2}}}}  \\
    t &= -\int_{r_0}^{r} dr \, \frac{e/f}{\sqrt{e^2- f(r)\lrp{1+\frac{\ell^2}{r^2}}}}  \; , \label{eq:t_nh_integration}
\end{align}
where $r_0$ is the radial distance of a fiducial point for the free fall of the atom consistent with the initial conditions in Eq. (\ref{eq:conserved-quantities}).  The integration of $\tau$ and $t$ can now be performed by using a Taylor expansion around the event horizon as a function of the near-horizon variable $x=r-r_+$. Up to first order in $x$, the integration yields,
\begin{align}
    \tau &= -\frac{x}{e}+\text{ const.} +  \mathcal{O}(x^2)\;,\label{eq:tau_in_x}\\
    t &= -\frac{1}{f_+'}\ln x - C x
    + \text{ const.} + \mathcal{O}(x^2)\;, \label{eq:t_in_x} 
\end{align}
where $C$ is a constant dependent on the conserved quantities given by
\begin{equation}
C = \frac{1}{2}\lrb{\frac{1}{e^2}\lrp{1+\frac{\ell^2}{r_+^2}}-\frac{f_+''}{(f_+')^2}} \; .
\end{equation}
Near the horizon we can neglect the $\mathcal{O}(x^2)$ terms as their contribution becomes negligible. This is equivalent to the hierarchical near-horizon expansion shown  in Eq. (\ref{eq:nh-expansions}). It should be noted that, while the coordinate time $t$ is logarithmic in $x$, the proper time $\tau$ is linear in $x$; in addition, the constant $C$ governs the linear term in the coordinate $t$. The logarithmic dependence on $x$ of the coordinate time ensures that it diverges when the particle reaches the horizon, i.e., when $x\rightarrow 0$, while the proper time $\tau$ remains finite.
  
Now we are equipped with all the quantities needed to calculate the excitation probability $P_{exc}$. Substituting Eqs.~(\ref{eq:conformal_field_modes}), (\ref{eq:tau_in_x}), and (\ref{eq:t_in_x}) into Eq.~(\ref{eq:P_ex_explicit}), we get
\begin{align}
    P_{exc} &= \frac{g^2}{e^2} \left|\int_0^{x_f} \; dx\;  x^{-i\Theta} e^{i\omega(-\ln x/f_+'-Cx)}e^{-i\nu x/e}\right|^2 \nonumber\\
    &= \frac{g^2}{e^2} \left|\int_0^{x_f} \; dx\;  x^{-i\sigma} e^{-i{s} x} \right|^2
    \label{eq:conformal_integral}
\end{align}
where
\begin{gather}
    \sigma = 2\Theta = \frac{2\omega}{f_+'} = \frac{\omega}{\kappa}\;,\label{eq:Theta_doubling} \\
    {s} = C\omega + \frac{\nu}{e} =  \frac{\omega}{2}\lrb{\frac{1}{e^2}\lrp{1+\frac{\ell^2}{r_+^2}}-\frac{f_+''}{(f_+')^2}} + \frac{\nu}{e}\;, \label{eq:frequency_definitions}
\end{gather}
and $\kappa={f_+'}/{2}$ is the surface gravity of the black hole. In Eq.~(\ref{eq:conformal_integral}), $x_f$ is an upper limit of the integration that signifies the boundary of a region where the near-horizon approximation is valid. Thus, Eq.~(\ref{eq:conformal_integral}) can be written in terms of the lower incomplete gamma function, but a conformal property of the integrand allows us to write the expression in a more familiar and compact form. As we will show in the next section, in the limit when ${s}\gg\sigma$, we can push the upper limit of the integration to infinity and evaluate the integral to give us 
\begin{align}
    P_{exc} & \approx \frac{2\pi g^2 \sigma}{e^2 {s}^2}\;\frac{1}{e^{2\pi\sigma}-1} \label{eq:planck_factor}   \\
 & =  \frac{2\pi g^2 }{\kappa} \, \frac{\omega}{\nu^2 \left( 1 + C e \, \omega/\nu \right)^2 } \; \frac{1}{e^{2\pi\sigma}-1} \label{eq:planck_factor_2}  \\
 & \approx  \frac{2\pi g^2 }{\kappa} \, \frac{\omega}{\nu^2} \; \frac{1}{e^{2\pi\omega/\kappa}-1}\; , \label{eq:planck_factor_3} 
\end{align}
where the approximation  $\nu\gg\omega$ is enforced again in the last step. This is the familiar Planck distribution with the Hawking temperature 
\begin{equation}
T=\frac{f_+'}{4\pi} = \frac{\kappa}{2\pi}
\;.
\end{equation}

The following remarks are in order. First, the frequency hierarchy $\nu \gg \omega$ is a ``geometrical optics'' approximation for the fall of the atom, i.e., a semiclassical treatment of the particle geodesics $\left( t(\tau), r (\tau) \right)$ as well-defined classical paths, which is necessary for this approach to be consistent. Second, applying this frequency hierarchy $\nu \gg \omega$ to $P_{exc}$ in the transition from Eq.~({\ref{eq:planck_factor_2}) to Eq.~({\ref{eq:planck_factor_3}) involves the reduction of the denominator $(1 + C e \,\omega/\nu)^{2} \approx 1$; this shows that the final result~({\ref{eq:planck_factor_2}) for $P_{exc}$ is independent of the numerical factor $C$. Third, $C$ is typically a numerical factor of order one that may also depend on the chosen rescaling of the near-horizon variable (see Appendix~\ref{app:reparametrization})---but this ``geometrical optics'' hierarchy removes any ambiguities in the selection of coordinates. Fourth, and most importantly, the derivation above shows that the final expression for $P_{exc}$ (with the removal of extra field-frequency factors), fully conforms to the Planck distribution at the Hawking temperature, i.e., it exhibits a similar behavior to the Hawking radiation itself.

\section{Conformal aspects of the near-horizon radiation\label{sec:conformal_aspects}}
\noindent In this section we provide a brief discussion on the conformal aspect of the radiation detected by the atom. From Eq.~(\ref{eq:conformal_integral}), the integrand consists of two multiplicative functions $f_1(x)\equiv x^{-i\sigma} =e^{-i \sigma\ln x}$ and $f_2(x)\equiv e^{-i{s} x} = e^{-i ( {s}/f_+' )( f_+' x ) }$, which are both oscillatory in nature. However, all aspects of the near-horizon physics, including the free-fall radiation properties under study, rely on the function $f_{1}(x)$, as will be proved below.

The oscillatory nature of the function $f_{1}(x)$ in Eq.~(\ref{eq:conformal_integral}) involves a spatial frequency that increases as $x \rightarrow 0$, i.e., as the event horizon is approached. This property arises from the logarithmic form of the phase of the near-horizon modes~(\ref{eq:conformal_field_modes}) of the governing CQM. Conformal invariance is manifested by the remarkable scaling symmetry of the modes. This invariance implies that the ensuing geometric pattern, displayed in Fig.~\ref{fig:Russian-doll-wavefronts}, looks identical under arbitrary magnifications. 
\begin{figure}[!h]
	\centering
	\includegraphics[width=0.6\linewidth]{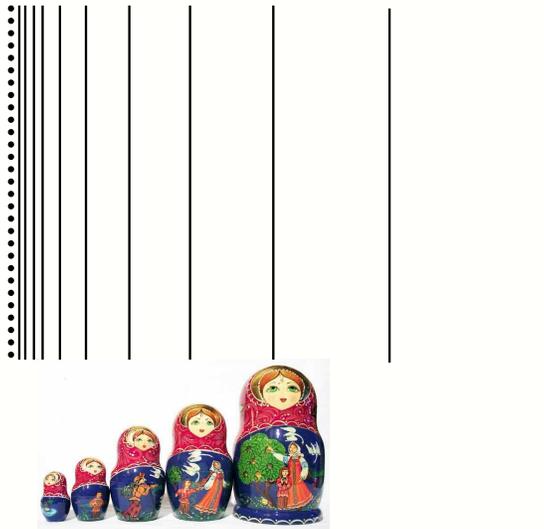}
	\vspace{-0.2in}
	\caption{The wavefronts associated with the near-horizon conformal modes  $\phi (r,t)$ are shown. The dotted line shows the location of the event horizon. The crowding of the modes as the horizon is approached follows from the sequence $x_{(n)} \propto \eta^{n}$. In this graph, we used an ad hoc value $\eta^{-1} = 1.4$. In general, $\eta = e^{-2\pi/\Theta}$, with $\Theta$ defined in Eq.~(\ref{eq:conformal_interaction}). The geometric scaling is depicted with the Russian-doll analogy. The function $f_1(x)$ exhibits an identical functional form, but with a doubling of the frequency scale, i.e., $\eta = e^{-2\pi/\sigma} = e^{-\pi/\Theta}$, as follows from Eqs.~(\ref{eq:conformal_integral}) and (\ref{eq:Theta_doubling}).   }
	\label{fig:Russian-doll-wavefronts}
\end{figure}
More precisely, {\em the pattern looks like a properly rescaled version of itself from any vantage point, i.e., invariant under rescaling transformations\/}. The meaning of this statement can be spelled out by identifying the functional form of the wavefronts associated with the modes. A given phase value for the mode $\phi (r,t)$ is achieved at coordinate values $x_{(n)}$ such that $\Theta \, \ln x_{(n)} = -2\pi n + \text{const}$, for integer $n$ (with the sign chosen so that, when $n \rightarrow \infty$, the event horizon is approached ). Therefore, $x_{(n)} = x_{(0)} \, \eta^{n}$ (defined up to a multiplicative constant) follows a geometric series with ratio $\eta = e^{-2\pi/\Theta}$. A complete characterization or equivalence of this geometric sequence is given by the relation 
\begin{equation}
\frac{ x_{(n'+m)} }{ x_{(n+m)} }= \frac{ x_{(n')} }{ x_{(n)}} 
\; .
\label{eq:geometric-scaling-ratio}
\end{equation}
(This equation is a straightforward corollary of $x_{(n)} = x_{(0)} \, \eta^{n}$, but it can also be iteratively reversed to reconstruct the whole geometric sequence.) In Eq.~(\ref{eq:geometric-scaling-ratio}), the vantage point is shifted from $n$ to $n+m$, thus proving the anticipated statement. Moreover, when $n \rightarrow \infty$, the geometric pattern of wavefronts exhibits infinite crowding towards the horizon as accumulation point, as shown in Fig.~\ref{fig:Russian-doll-wavefronts}. These properties have been pointed out by several authors---most notably in Refs.~\cite{DeWitt_QFT-curved} and~\cite{Jacobson_QFT-curved}. But it is only by highlighting the governing role of CQM that such behavior and its universal manifestations for thermal radiation and black hole thermodynamics can be fully understood. Incidentally, other aspects of this ``Russian-doll'' behavior have been studied in CQM in terms of renormalization frameworks and a variety of physical realizations~\cite{LeClair_Russian-doll,Glazek_limit-cycles,HEC-CRO_CQM}. 

It is noteworthy that the functional dependences of the modes $\phi^{*} (r,t)$ and the function $f_1(x)$ are equivalent because $f_1(x)$ arises from $\phi^{*} (r,t)$ in Eq.~(\ref{eq:P_ex_explicit}) in the near-horizon limit, 
\begin{equation}
 \phi^{*} (r,t(\tau(r))) \stackrel{(\mathcal H)}{\sim} f_1(x) \, e^{-i C \omega x}
 \label{eq:f_1-from-phi*} 
 \; ,
 \end{equation} 
where $x= r -r_{+}$ should be used on the right-hand side, and the extra factor $e^{-i C \omega x}$ appears at higher orders and can be neglected, as discussed at the end of Sec.~\ref{sec:nhqoapproach}}. Specifically, there is an additional logarithmic $x$ dependence of the coordinate time $t$ via the proper time $\tau$ of the atom's geodesic in Eq.~(\ref{eq:f_1-from-phi*}). Thus, while the explicit $x$ dependence of the modes is
 $\phi^{*} (r,t) \stackrel{(\mathcal H)}{\sim} x^{-i \Theta} e^{i \omega t} \propto x^{-i \Theta} 
 $ according to Eq.~(\ref{eq:conformal_field_modes}), the total $x$ dependence of $f_1(x)$ becomes 
 $f_1(x) \stackrel{(\mathcal H)}{\sim} x^{-i \sigma}= x^{-2 i \Theta }$, which involves the doubling of the scale $\Theta \rightarrow 2 \Theta$. The former probes the field in static Schwarzschild coordinates  while the latter probes the field following the freely falling atom. But the patterns associated with both functions have the same geometric form shown in Fig.~\ref{fig:Russian-doll-wavefronts}.
 
In this paper, the conformal behavior driven by the function $f_1(x)$ leads directly to the Planck distribution, with the Hawking temperature, of the radiation emitted by the freely falling atom. This can be seen from the integral of Eq.~(\ref{eq:conformal_integral}), where the upper limit of integration is $x_f \ll r_+$, such that the near-horizon approximation is valid. As we will show below, from the conformal behavior of $f_{1}(x)$, we can push the upper limit to infinity without significant error, and thus derive the Planck distribution. The validity of this approximation is due to the nontrivial $x$-dependent frequency resolution of $f_1(x)$. The same oscillatory logarithmic dependence of the wave-like function $f_{1}(x)$---which produces an increasing, diverging spatial frequency near the event horizon---leads to decreasingly slower variations with respect to $x$ much farther away, i.e., for $f_+' x\geq \mathcal{O}(1)$ (Fig.~\ref{fig:Russian-doll-graphs}(b)). It should be noted that the factor $f_+' $ provides a characteristic inverse length scale (of the order or $1/r_+$), which permits a comparison of the various other parameters involved. 

\begin{figure}[!h]
	\centering
	\includegraphics[width=0.9\linewidth]{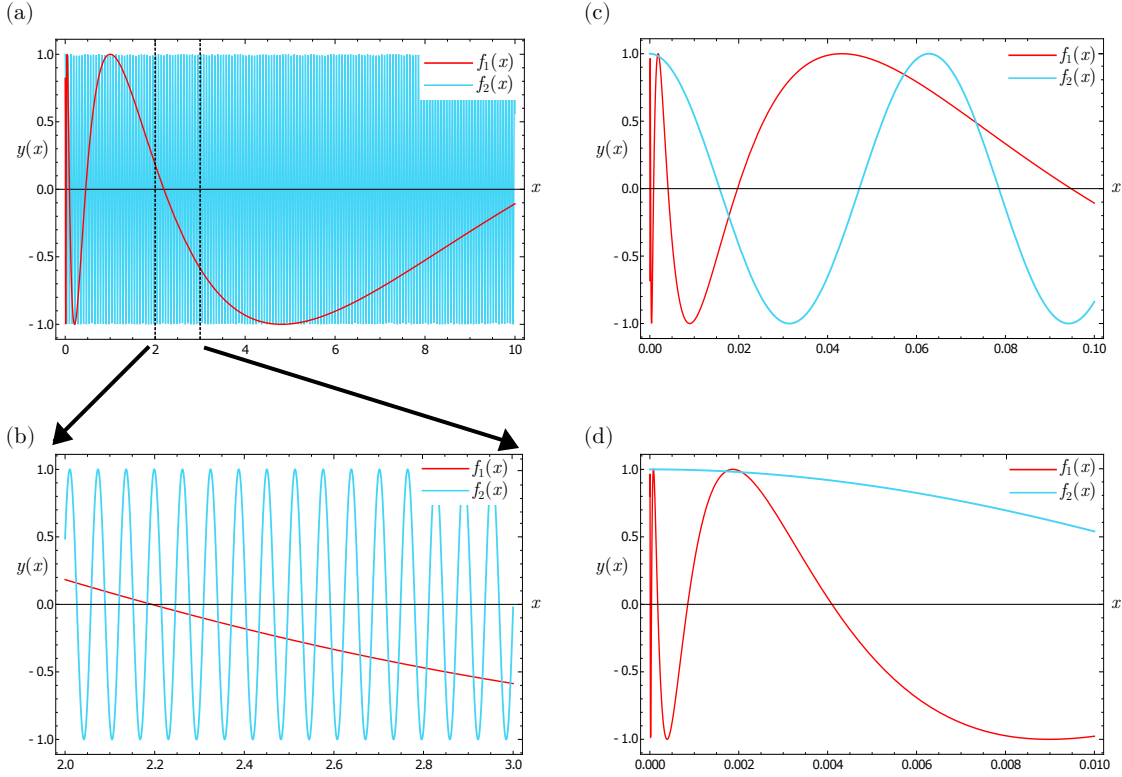}
	\caption{The graphs depict the oscillatory functions $f_1(x)$ and $f_2(x)$. The conformal nature of $f_1(x)$ matches the Russian-doll behavior shown in Fig.~\ref{fig:Russian-doll-wavefronts}, with a local spatial frequency that can be compared against the single-frequency oscillating function $f_2(x)$. For the sake of simplicity, only the real part of the functions $f_1(x)$ (shown in red) and $f_2(x)$ (in blue) are plotted. The imaginary parts behave in a similar way. The parameters used to generate the plots are $\sigma = 2.0$ and ${s} = 50.0$. (a) Plot of $\Re[f_1(x)]$ and $\Re[f_2(x)]$ in a range of 0 to 10. (b) A magnified view of a region in (a) to show the slow variation of $f_1(x)$ compared to $f_2(x)$. (c) and (d) show that, as we zoom in closer to the origin, the behavior of $f_1(x)$ remains scale invariant, whereas the oscillation becomes comparable to that of $f_2(x)$.}
	\label{fig:Russian-doll-graphs}
\end{figure}

In contrast with the nontrivial frequency resolution of $f_1(x)$, the competing oscillating function $f_2(x)$ is controlled by a single spatial frequency $s$. Then, for ${s}/f_+' \gg \sigma$ and $ f_+' x\geq \mathcal{O}(1)$, $f_2(x)$ oscillates very rapidly compared to the relatively slower changes of $f_1(x)$, as shown in Fig.~\ref{fig:Russian-doll-graphs}(a,b). Thus, the contribution of the integrand becomes negligible on average. Furthermore, the condition  ${s}/f_+' \gg \sigma$ is again essentially due to the ``geometrical optics'' frequency hierarchy $\nu \gg \omega$. On the other hand, as we move closer to the origin, the variations of $f_1(x)$ and $f_2(x)$ become comparable with changing $x$ (Fig.~\ref{fig:Russian-doll-graphs}(c)). It is this region that contributes significantly to the sum. Moving closer to the event horizon (``origin'' $x=0$ for the variable $x$) reveals the Russian-doll behavior described above for the function $f_{1}(x)$; however, the function $f_2(x)$ becomes slowly varying on that near-horizon scale (Fig.~\ref{fig:Russian-doll-graphs}(d)). This indicates that only the near-horizon region contributes significantly to the integral in Eq.~(\ref{eq:conformal_integral}); thus, extending the limit of integration does not significantly affect the value of the integral due to the rapid oscillations of the function $f_2(x)$. 

As a consequence, the excitation probability can be evaluated by the following sequence of steps:
\begin{equation}
    P_{exc} = \frac{g^2}{e^2} \left|\int_0^{x_f} \; dx\;  x^{-i\sigma} e^{-i{s} x} \right|^2  \longrightarrow \frac{g^2}{e^2} \left|\int_0^{\infty} \; dx\;  x^{-i\sigma} e^{-i{s} x} \right|^2 = \frac{2\pi g^2 \sigma}{e^2{s}^2}\;\frac{1}{e^{2\pi\sigma}-1}\;.
    \label{eq:P_ex_steps} 
\end{equation}

In summary, in this section we have outlined a heuristic argument that explains why the extension of the upper limit of integration is asymptotically valid when ${s}/f_+' \gg \nu$. This procedure can be further justified by a more rigorous, analytical approach, as discussed in Appendix~\ref{app:incomplete_gamma}, where the explicit evaluation of the integral in the last step of Eq.~(\ref{eq:P_ex_steps}) is also spelled out.

\section{Discussion \label{sec:Discussion}}
\noindent 
We have developed a conformal approach to provide a deeper understanding of the nature of the radiation emitted by an accelerated atom, and its relationship with the Hawking effect. This conformal approach involves a reexamination of the model advanced in Ref.~\cite{scully2018}, where it was shown that an atom falling radially towards a Schwarzschild black hole, with zero kinetic energy at infinity, experiences radiation in the Boulware vacuum. This acceleration radiation is ultimately due to the relative acceleration between the atom and the outgoing photon modes. Moreover, the specific realization of this radiation---via transitions associated with counterrotating terms in the interaction Hamiltonian~(\ref{QO_interaction_potential})---further supports the existence of Unruh radiation, despite claims to the contrary~\cite{Unruh_contrarians}.

In this article, in addition to displaying the near-horizon aspects of the acceleration radiation by free fall as arising from conformal quantum mechanics, we have extended the applicability of this model to the broader class of generalized D-dimensional Schwarzschild metrics with general initial conditions. The atom undergoes a free fall with conserved total specific energy $e$ and angular momentum $\ell$ (per unit mass). Carrying out the detection probability calculation is nontrivial and more involved in the original formalism, and limited to the particular initial conditions defined by $e=1$ and $\ell =0$. However, exploiting the conformal nature of the physics near the horizon provided us with a much simpler method. Using the near-horizon expansion for the field modes and the geodesic equations, we have shown that the probability distribution can be derived in a closed form. Furthermore, in the ``geometrical optics'' limit  $\nu\gg\omega$, which implies ${s}\gg\sigma$, the probability distribution reduces to the Planck form with the Hawking temperature. The conformal nature of the integral in Eq.~(\ref{eq:conformal_integral}) plays a crucial role in obtaining the Planck distribution, as discussed in Sec.~\ref{sec:conformal_aspects}. In Appendix~\ref{app:reparametrization}, we also show that the result is robust under a reparametrization of the near-horizon variable within the same approximation $\nu\gg\omega$.

In summary, the formalism developed here provides further insight into the role played by the event horizon and the near-horizon physics in the acceleration radiation by free fall. Moreover, the techniques we have presented in this paper also allow us to handle more general spacetime geometries and initial conditions. Extensions of this work are in progress, and include exploring more general classes of spacetime backgrounds (e.g., including black hole rotation) and finding the deeper connection between singular CQM and the Unruh effect.

\acknowledgments{}
This research was supported in part by the University of San Francisco Faculty Development Fund (H.E.C.). One of the authors (C.R.O.) would like to thank Dr.\ M. O. Scully for the stimulating conversations that prompted this investigation.

\begin{appendix}

\section{Derivation of CQM equation from Klein-Gordon equation \label{app:cqm}}	
\noindent The Klein-Gordon equation (\ref{eq:Klein_Gordon_basic}), with the choice of metric given in Eq.~(\ref{eq:RN_metric}), reduces to
\begin{equation}
- \frac{1}{f} \ddot{ \Phi}+ f \Phi'' + \left(f' + \frac{(D-2) f}{r} \right)\Phi'+ \frac{1}{r^{2}} 
{\Delta}^{(D-2)}_{(\gamma)} \Phi - \left(m^{2} + \xi R\right)\Phi = 0\; ,
\label{eq:Klein_Gordon_radial_metric}
\end{equation} 
where the dots and primes stand for time and radial derivatives respectively, while ${\Delta}^{(D-2)}_{(\gamma)} $ and $\gamma_{ab}(\Omega)$ are the Laplacian and the metric on $S^{D-2}$. The quantum field operator can be expanded as
\begin{equation}
\Phi(t,r,\Omega) = \sum_{n,l,m} \lrb{a_{nlm} \, \phi_{nlm}(r,\Omega,t) + H.c.}. \label{eq:quantizationappendix}
\end{equation}
with the following ansatz as mentioned in the main text Eq.~(\ref{eq:separation_of_variables})
\begin{equation}
\phi_{nlm} (t,r, \Omega) = \chi (r)\, u_{nl}(r)Y_{lm}(\Omega)e^{-i\omega_{nl}t}\;.
\end{equation}
The choice of the radial function
\begin{equation}
\chi (r)= \exp \left\{ -\frac{1}{2} \int \left[ \frac{f'}{f} + \frac{(D-2)}{r} \right] dr \right\}=[f(r)]^{-1/2} \,r^{-(D-2)/2}\; .
\label{eq:transf_factor_Liouville}
\end{equation}
brings the radial part of Eq.~(\ref{eq:Klein_Gordon_radial_metric}) to the canonical form
\begin{equation}
u_{nl}''(r) +I_{(D)} (r; \omega_{nl}, \alpha_{l,D}  )\,u_{nl}(r) =0\;  ,
\label{eq:Klein_Gordon_normal_radial_app}
\end{equation}
where
\begin{eqnarray}
I_{(D)}  (r; \omega, \alpha_{l,D}  ) 
& = &
\frac{ 1 }{f^{2}}\left( \omega^{2} + \frac{f'^{2}}{4}\right)-\frac{\left(  m^{2} + \xi R \right) }{f} 
-\frac{1}{f}\frac{\alpha_{l,D} }{ r^{2} }  \nonumber\\
&  & 
+ R_{rr} +\left\{\left(\frac{1}{f} - 1\right) \left[(D-3)/2\right]^{2}+ \frac{1}{4}\right\} \frac{1}{r^{2}} \;  ,
\label{eq:Klein_Gordon_normal_radial_invariant}
\end{eqnarray}
with 
\begin{equation}
R_{rr}=-\frac{f''}{2 f}-\frac{(D-2)}{r} \, \frac{ f'}{2  f} \;  
\label{eq:Ricci_rr}
\end{equation}
being the radial component of the Ricci tensor for the metric~(\ref{eq:RN_metric}) and
\begin{equation}
\alpha_{l,D} =l(l+D-3) + \left[(D-3)/2\right]^{2} = \left(l + \frac{D-3}{2} \right)^{2}
\end{equation}
being the angular momentum coupling. \\
Now, with the near-horizon expansion described in Eq.~(\ref{eq:nh-expansions}), the various terms in Eq.~(\ref{eq:Klein_Gordon_normal_radial_invariant}) can be reduced with the replacements $f''/f \stackrel{(\mathcal H)}{\sim} f''_{+}/(f'_{+}x) $ and $f'/f \stackrel{(\mathcal H)}{\sim}  1/x$, together with 
$r \stackrel{(\mathcal H)}{\sim} r_{+}$; in each one of these factors, the corrections are multiplicative and of the order $[1+ \mathcal{O}(x)]$. As a result, the leading orders of each one of the terms on the right-hand side of Eq.~(\ref{eq:Klein_Gordon_normal_radial_invariant}) become
\begin{eqnarray}
\! \! \! \! &  &I_{(D)} (r; \omega, \alpha_{l,D}  ) \stackrel{(\mathcal H)}{\sim}\frac{ 1 }{(f'_{+})^{2}}\left[ \omega^{2} + 
\frac{(f_{+}')^{2}}{4}\right]\frac{1}{x^{2} }\,[ 1 + \mathcal{O}(x) ]-\frac{\left(  m^{2} + \xi R_{+} \right) }{f'_{+}}
\frac{1}{x}\,[ 1 + \mathcal{O}(x) ]\nonumber \\
\! \! \! \! & - & \! \! \frac{1}{f'_{+}}\frac{\alpha_{l,D} }{ r_{+}^{2} }\frac{1}{x}\,[ 1 + \mathcal{O}(x) ]  - \left[\frac{ f_{+}''}{ 2 f'_{+}} + \left( \frac{D}{2} -1 \right) \frac{1}{ r_{+}}\right] \frac{1}{x}\,[1 + \mathcal{O}(x) ]
+\frac{1}{f'_{+}} \frac{1}{r_{+}^{2}} \frac{1}{x}\,[ 1 + \mathcal{O}(x) ]\, ,
\label{eq:Klein_Gordon_normal_radial_invariant_leading_orders}
\end{eqnarray}
and the {\em leading\/} term in Eq.~(\ref{eq:Klein_Gordon_normal_radial_invariant_leading_orders}), of order $\mathcal{O}(1/x^{2})$, becomes asymptotically dominant as $r \stackrel{(\mathcal H)}{\sim} r_{+}$.
Therefore, Eq.~(\ref{eq:Klein_Gordon_normal_radial_invariant}) yields the CQM equation
\begin{equation}
u''(x)+\frac{ \lambda_{\rm eff} }{x^{2}}\,\left[ 1 + \mathcal{O}(x) \right]u (x)=0\;  ,
\label{eq:Klein_Gordon_conformal_app}
\end{equation} 
where we have written $u(r)\equiv u(x)$, and 
\begin{equation}
\lambda_{\rm eff} = \frac{1}{4} + \Theta^{2}\, , \; \; \; \; \Theta= \frac{\omega}{  f'_{+} } \; .
\label{eq:conformal_interaction_app}
\end{equation}  
The validity of the expansion in $x$ relies on the condition $r - r_{+} = x \ll r_{+}$.

\section{Equivalence of Eddington-Finkelstein modes and CQM modes \label{app:efconformal}}
\noindent In  Eddington-Finkelstein coordinates, the leading order of the outgoing field mode is proportional to
\begin{equation}
    \phi(r,t) = e^{-i\omega(t-r_*)}\;.
\end{equation}
where $r_*$ is the tortoise coordinate given by the integral
\begin{equation}
    r_* = \int \frac{dr}{f(r)} \stackh \int \frac{dx}{f_+'x}\lrp{1-\frac{f_+''}{f_+'}\frac{x}{2}} = \frac{1}{f_+'}\ln x -  \frac{f_+''}{2(f_+')^2}x\;.
\label{eq:tortoise-nh}
\end{equation}
Thus, enforcing the near-horizon approximation~(\ref{eq:tortoise-nh}), the Eddington-Finkelstein field modes can be written as
\begin{equation}
    \phi(r,t) = \exp ({-i\omega t}) \,
    \exp \left( {i\frac{\omega}{f_+'}\ln x-i\frac{\omega f_+''^2}{2(f_+')^2}}x \right) = e^{-i\omega t}x^{i\Theta} \,
    \exp \left({-i\omega\frac{f_+''^2}{2(f_+')^2}x} \right) \;.
    \label{eq:eq:EF-modes}
\end{equation}
Compared to the leading CQM modes of Eq.~(\ref{eq:conformal_field_modes}), we see that Eq.~(\ref{eq:eq:EF-modes}) has the same governing exponential factors $e^{-i\omega t}$ and $x^{i\Theta}$. In addition, the extra factor in Eq.~(\ref{eq:eq:EF-modes}), appears from the next-to-leading order near-horizon approximation of these modes, arising from the second term in Eq.~(\ref{eq:tortoise-nh}). This extra factor, when used in the evaluation of  $P_{exc}$ in Eq.~(\ref{eq:P_ex_explicit}), modifies the constant $C$ defined in Eq.~(\ref{eq:frequency_definitions}), which now becomes
\begin{equation}
C_{\rm EF} = C-\frac{1}{2} \frac{f_+''}{(f_+')^2} =
\frac{1}{2 e^2} \lrp{1+\frac{\ell^2}{r_+^2}}-\frac{f_+''}{(f_+')^2} \; .
 \end{equation}
However, in the ``geometrical optics'' limit $\nu\gg\omega$ (see the last paragraph of Sec.~\ref{sec:nhqoapproach}), the value of this constant is not relevant, Moreover, this allows us to write the excitation probability as a Planck distribution regardless of the modes selected at intermediate steps. 

\section{Conformal aspects of the integral in Eq.~(\ref{eq:conformal_integral})\label{app:incomplete_gamma}}
\noindent In this appendix we investigate the mathematical structure of the integral in Eq.~(\ref{eq:conformal_integral}) in greater detail. We begin by recasting the integral into a form of gamma function as shown below. Specifically,
\begin{align}
    \mathcal{\hat{I}}=  \int_0^{x_f} dx\; e^{ -i x {s} } x^{-i\sigma} = \frac{1}{(i{s})^{1-i\sigma}}\int_0^{y_f} dy\;y^{-i\sigma}e^{-y} = \frac{\gamma(1-i\sigma,y_f)}{(i{s})^{1-i\sigma}}\;, \label{eq:incomplete_gamma}
\end{align}
where $\gamma(z,b)$ is the lower incomplete gamma function defined by
\begin{equation}
    \gamma(z,b) = \int_{0}^{b} dy\; e^{-y} y^{z-1} \;.
\end{equation}
We claim that, in the approximation ${{s}}\gg\sigma$, the upper limit can be pushed to infinity and the integral can be written as
\begin{equation}
   \mathcal{\hat{I}} = \frac{\Gamma(1-i\sigma)}{(i{s})^{1-i\sigma}}\;.
\end{equation}
Here, $\gamma(z,\infty) = \Gamma(z)$ is the ordinary gamma function when the upper limit of the integral is infinity. This integral can be further rewritten as
\begin{equation}
 \mathcal{\hat{I}}
=
\frac{ | \Gamma ( 1- i \sigma) |  \, e^{-\pi \sigma/2} }{
 {s}  } \, e^{i \delta} 
 =
 {s}^{-1}
  \sqrt{ \frac{ 2 \pi \sigma }{  e^{2 \pi \sigma} -1 }}
    \; e^{i \delta} 
    \; ,
\end{equation}
where use was made of $ \displaystyle | \Gamma ( 1-  i \sigma) | \,  e^{-\pi \sigma/2} = \sqrt{  \frac{2 \pi \sigma}{ e^{2 \pi \sigma} -1 }} $, and $\delta$ is a real phase. Therefore, the relevant probability factor in Eqs.~(\ref{eq:P_ex_explicit}) and (\ref{eq:planck_factor}) becomes
\begin{equation}
|\mathcal{\hat{I}}|^{2}
=
\frac{ 2 \pi \sigma }{ {s}^2 \left( e^{2 \pi \sigma} -1 \right)} 
\; .
\end{equation}
This leads to the final expressions for the probability integral,
\begin{equation}
    P_{exc} = \frac{g^2}{e^2 } \left| \frac{\Gamma(1-i\sigma)}{(i{s})^{1-i\sigma}}\right|^2 = \frac{2\pi g^2 \sigma}{e^2{s}^2}\;\frac{1}{e^{2\pi\sigma}-1}\;.
\end{equation}
Parenthetically, in calculating the analytic continuation of the gamma function to the complex plane, one should add a small real $\epsilon$ to the exponent of $y$ in the integral in Eq.~(\ref{eq:incomplete_gamma}) and then take the limit $\epsilon\rightarrow 0$ after the integral is carried out.

One can probe the properties of the incomplete gamma function to find the validity of the approximation of pushing the upper limit to infinity, but it is more instructive to investigate the form of the integral in terms of the variable $x$. The method of stationary phase implies that, for an integral of the form $\int f(x) e^{izg(x)} dx$, where $z\rightarrow \infty$ and $g(x)$ is bounded, the contribution comes from the boundary points and the stationary points of $g(x)$. Since there is no stationary point for the function $g(x)$ in the probability integral, the only contribution comes from the boundary points (see Ref.~\cite{belyaninetal} for a detailed discussion). Now, if we displace the upper boundary $x_f$ to $x_f+\delta x$, then the extra contribution to the integral from the interval $(x_f,x_f+\delta x)$ will be negligible because it does not contain any stationary point. In this way, we can push the upper boundary to infinity without adding any significant contribution to the integral, thus validating the approximation made in the main text. 

\section{Reparametrization of the near-horizon variable\label{app:reparametrization}}
\noindent In this appendix, we show that a reparametrization of the near-horizon variable can produce inequivalent results for the probability integral, but, in the ``geometrical optics'' limit $\nu\gg\omega$, all reparametrizations lead to the same prefactor in Eq.~(\ref{eq:planck_factor}). This is indeed the ``geometrical optics'' condition discussed at the end of Sec.~\ref{sec:nhqoapproach}. To simplify calculations we use the dimensionless variables 
$\xi=(r-r_+)/r_+$ and normalize $r_+$ to 1. This essentially enables us to replace $x$ with $\xi$. Thus, the near-horizon expansion is an expansion in $\xi$. In this new variable, we then consider the near-horizon class of reparametrizations of $\xi$ defined by
\begin{equation}
\xi = (1+ \eta)^{\alpha} - 1\; ,
\label{eq:reparametrization}
\end{equation}
which are labeled by the parameter $\alpha$, and where $\eta \equiv \tilde{\xi}$ is the redefined near-horizon variable. When we further enforce the near-horizon approximation, the leading orders of the variables are related by 
\begin{align}
    \xi &\stackh \alpha \eta  \label{eq:scale_transf} \\
    \ln \xi &\stackh \ln \eta + \frac{(\alpha-1) \, \eta}{2} + \text{ const.} \label{eq:reparam_log}
\end{align}
The expansions just change the constant ${s}$ in Eq.~(\ref{eq:frequency_definitions}) by an extra factor. This leads to superficially inequivalent results for Eq.~(\ref{eq:planck_factor}), where in the denominator includes a shifted value of ${s}$. However, if we consider the ``geometrical optics'' limit $\nu\gg\omega$, all the inequivalent expressions converge to the same result obtained in Ref.~\cite{scully2018}, i.e.,
\begin{equation}
    P_{exc} = \frac{2\pi g^2 \sigma}{\nu^2}\;\frac{1}{e^{2\pi\sigma}-1}\;.
\end{equation}
This proves the robustness of the radiation formula under this reparametrization. Moreover, this is a self-consistent result in that the approximation $\nu\gg\omega$ is also needed in the form $s\gg\sigma$ for the derivation of Eq.~(\ref{eq:planck_factor}), as outlined in Sec.~\ref{sec:conformal_aspects}.

In the near-horizon limit, the reparametrization procedure defined above differs from a pure scale transformation due to the presence of the logarithmic term in the expression for the field modes. For a scaling transformation of the near horizon variable $\xi\rightarrow \tilde{\xi} = \alpha\xi$, the result of Eq.~(\ref{eq:planck_factor}) remains invariant. This is not surprising because the near horizon expansion follows the CQM defining condition, Eq.~(\ref{eq:Klein_Gordon_conformal}), which is scale invariant. The difference between the reparametrization~(\ref{eq:reparametrization}) and a scale transformation~(\ref{eq:scale_transf}) is the expansion of the logarithmic function, where a parameter-dependent extra term appears, as shown in Eq.~(\ref{eq:reparam_log}). However, the parameter-dependent term becomes irrelevant when evaluating 
Eq.~(\ref{eq:P_ex_explicit}), in the hierarchical limit $\nu\gg\omega$. Thus, the reparametrization is equivalent to a scale transformation within the ``geometrical optics'' approximation---see the comments in the last paragraph of Sec.~\ref{sec:nhqoapproach}. With these qualifications, our results are in agreement with those of Ref.~\cite{scully2018}, by using a transformation of the type~(\ref{eq:reparametrization}) with $\alpha =2/3$.

In short, a reparametrization of the coordinate can lead to an inequivalent prefactor for the probability distribution. However, this issue can be resolved by considering the limit $\nu\gg\omega$, where all inequivalent expressions of $P_{exc}$ converge to the unique thermal Planck distribution, thus displaying a robustness of the result under this transformation. 

\end{appendix}

\end{document}